\documentstyle[12pt,a4]{article}

\begin{document}

\addtolength{\baselineskip}{0.2 \baselineskip}

\begin{center}
{\bf \LARGE Absorption mechanisms in photon induced two-body knockout}
\\
\vspace{1.cm}

{\Large L. Machenil \footnote[2]{Research fellow IWONL},
M. Vanderhaeghen \footnote[1]{Research fellow NFWO},
J. Ryckebusch \footnotemark[1]
\\ and M. Waroquier \footnote[3]{Research Director NFWO}}
\\
\vspace{1.5 cm}

{\em Laboratory for Nuclear Physics and Laboratory for Theoretical
Physics, \\ Proeftuinstraat 86, 9000 Gent, Belgium}
\end{center}

\begin{abstract}
Calculations have been performed for the $^{16}$O($\gamma$,pn) and the
$^{16}$O($\gamma$,pp) reaction in the photon-energy range $E_{\gamma}$ =
60-300 MeV. Besides the contribution from the more common
photoabsorption on the pionic degrees of freedom, we have investigated
the influence of heavier meson exchange ($\rho, \sigma, \omega$) and
intermediate $\Delta$ creation with $\pi$ and $\rho$ exchange. Whereas
the $\pi$ meson is found to set the main trends, the $\rho$ meson is
found not to be discardable in a theoretical description of the
($\gamma$,pn) reaction. The incorporation of an energy dependence and a
decay width in the $\Delta$ propagator is observed to be essential in order
to arrive at a more realistic description of ($\gamma$,NN) reactions at
higher photon energies.
\end{abstract}

\section{Introduction}

Photo-induced two-nucleon emission reactions are generally considered as
a valuable tool for the investigation of the two-body aspects of nuclear
structure \cite{Got58,Gar81}. Since the early eighties the major
contribution to the ($\gamma$,pn) strength is supposed to come from
$\pi$-meson exchange currents ($\pi$-MEC)
\cite{Wak83,Boa89,Bof91,Ryc92}. As these currents involve charge
exchange they do not contribute to the ($\gamma$,pp) reaction. In a direct
knockout reaction mechanism after photoabsorption on a two-body nuclear
current, the ($\gamma$,pp) channel can be fed through intermediate
$\Delta$(1232) isobar creation and through heavier meson exchange.  In
addition to this, the short range correlations (SRC) were suggested
\cite{Giu92} to contribute considerably into this channel. The pn
channel being considerably stronger, rescattering into the pp channel
could be expected to give important contributions to ($\gamma, pp$) an
effect which still needs to be investigated.  Continuous efforts to
improve on the experimental facilities have resulted in a renewed
interest in the field of photon induced two-nucleon emission processes.
In the near future, high-resolution ($\gamma$,pn) and ($\gamma$,pp) data
can be expected from Lund, Mainz and Saskatchewan. The progress made in
the construction of detection systems should also make the measurement
of angular ($\gamma$,NN) cross sections feasible in the next coming
years \cite{McGregor}. These developments have prompted us to study the
physics of the photoabsorption mechanism in ($\gamma, pn$) and
($\gamma$,pp) reactions in more detail. At this stage all theoretical
studies have employed pionic degrees of freedom only. In this work we
will extend this approach and in addition to the pion, we will study the
influence of heavier meson ($ \rho, \sigma, \omega$) exchange on the
($\gamma$,NN) cross section.  Moreover, we will investigate under which
kinematical conditions effects due to heavier meson exchange can best be
looked for.

\section{The formalism}

In this letter we will focus on the different absorption mechanisms
playing a role in the two-nucleon knockout processes. In order to
explore this topic in the most transparant conditions we will use a
model in which the nuclear structure effects are clearly separated from
the physics of the photoabsorption. Any model that aims at treating the
final state interaction (FSI) between the outgoing particles and the
residual nucleus starting from principal grounds is unfactorized in
nature. In such a model the physics of  the photoabsorption mechanism
and the nuclear-structure aspects of the reaction cannot be separated.
Therefore we do not include FSI between the outgoing particles and the
residual nucleus nor between the outgoing particles themselves. In a
previous paper \cite{Ryc92} we have shown that the Gottfried approach is
a good research tool as it is numerically tractable and gives a fair
approximation to the cross section. In the Gottfried approach,  the
differential cross section factorizes in a dynamical part $S_{fi}$ which
reflects the absorption mechanism and in the probability
$F_{hh'}(\bf{P})$ to find a NN pair in the single-particle orbits $h$ and
$h'$ with total momentum $\bf{P}$ and zero separation in the target
nucleus :
\begin{equation}
\frac{d^5\sigma^{lab}} {d\Omega_a d\Omega_b dp_b} \sim F_{hh'} S_{fi}.
\end{equation}
In order to calculate $S_{fi}$, a microscopic model for the two-body
nuclear currents is needed. For the calculations presented here, we
account for $\pi, \rho, \sigma$ and $\omega$ meson exchange currents and
$\Delta$ isobar currents. The meson-nucleon and meson-isobar coupling
constants are taken from a one-boson-exchange parameterization of the
Bonn potential \cite{bonn}.

The expression of the pionic exchange currents has been derived from the
chiral Lagrangian from ref. \cite{peccei}. For pseudovector $\pi$NN
coupling, this leads in lowest order to the $\pi$ seagull current
(fig.~1a) and the $\pi$ in-flight current
(fig.~1b). The combined $\pi$ seagull and $\pi$ in-flight
current satisfy the continuity equation with the one pion exchange
potential (OPEP). We used the nonrelativistic reduction of these
currents as they can e.g. be found in  ref.\cite{riska}. Due to their
isospin structure, these currents can only contribute to pn emission in
a direct knockout model. For the $\pi NN$ vertex, a monopole form factor
with cut-off mass 1200 MeV \cite{bonn} has been  used. The use of
hadronic form factors to regularize the $\pi$NN vertices at short
internucleon distances,  makes the introduction of additional currents
necessary in order to conserve the gauge invariance conditon with OPEP
\cite{mathiot}. These  contributions have been taken into account but
were found to be small for the photon energies considered in this paper.

The $\rho$ exchange currents have next to a $\rho$ seagull (fig.~1a) and
a $\rho$ in-flight contribution (fig.~1b), also a $\rho$ pair
contribution (fig.~1c) in which the intermediate nucleon propagator only
contains the antinucleon contribution. To derive these $\rho$ exchange
current contributions consistently into order ($1/M^2$) (with  M the
nucleon mass), it is necessary to start with a relativistic description
of the $\rho NN$ vertex \cite{bonn}.The $\rho$ exchange diagrams are
then calculated fully relativistically. The nonrelativistic reduction
(keep all terms to order ($1/M^2$)) of this expression yields the $\rho$
exchange currents which were used in this paper. It was checked that
their expressions agree with the ones given in ref. \cite{towner}. A
systematic discussion of the influence of the various $\rho$ exchange
current contributions as function of the photon energy will be given in
a forthcoming paper. In this paper we will study the importance of the
total $\rho$ exchange current as compared to the $\pi$ exchange current.
For the $\rho NN$ vertex, we also use a monopole hadronic form factor
with a cut-off mass 1500 MeV.

Being neutral mesons, the $\sigma$ and $\omega$ mesons can only
contribute to the current through the pair diagram (fig.~1c). The
expression for the $\sigma$ and $\omega$ exchange currents is obtained
along the same lines as described for the $\rho$ meson.

In deriving the current related to intermediate $\Delta$ excitation we
included both pion and rho exchange (fig.~1d) and adopted a
nonrelativistic approach. This means that $\pi N\Delta$, $\rho N\Delta$
and $\gamma N\Delta$ vertices are obtained from $\pi NN$, $\rho NN$ and
$\gamma NN$ vertices by replacing spin and isospin operators by the
corresponding transition operators. In the energy denominator which
describes the nonrelativistic delta propagation, the (photon) energy
dependence is retained. To account for the finite lifetime of the delta,
an energy dependent width is introduced which satisfies unitarity
\cite{width} in the $\pi N$ channel. The hadronic form factors at the
$\pi N\Delta$ and $\rho N\Delta$ vertices are taken to be the ones at
the $\pi NN$ and $\rho NN$ vertices.   The $\Delta$ currents with rho
exchange are noticed to  contain as a subset the terms of the $\Delta$
current with pion exchange but with opposite sign. Because of the strong
$\rho NN$ coupling compared to the $\pi NN$ coupling a strong
destructive interference between both contributions could be expected.

\section{Results}

All forthcoming cross sections were obtained for the
$^{16}$O($\gamma$,NN) reaction in planar kinematics (i.e. the two
emitted nucleons and the photon remain in the same fixed plane), except
when stated explicitly. The cross sections involve two nucleon emission
from all occupied sp orbits, where the hole states are described with
harmonic oscillator wave functions.

To start with we concentrate on the ($\gamma$,pn) reaction. In fig.~2 we
displayed the contribution from the different absorption mechanisms to
the integrated $^{16}$O($\gamma$,pn) cross section as a function of the
photon energy. We have calculated the contribution arising from the two
low mass isovector mesons ($\pi$ and $\rho$) which give rise to the
electromagnetic exchange operators through the seagull and the in-flight
diagrams (fig.~1a-b) and also through the pair diagram (fig.~1c) for the
$\rho$ meson. The exchange current contributions from the two isoscalar
mesons ($\sigma$ and $\omega$) were found to be small, because they only
contribute through the pair diagram (fig.~1c) which is of relativistic
origin. Even at low photon energies the interference between $\pi$ and
$\rho$ meson exchange reveals an appreciable reduction of the cross
section (fig.~2), indicating the importance to incorporate $\rho$ meson
exchange in the reaction process.

It is tempting to simulate the reducing effect of the $\rho$ meson on
the total strength, by using an effective cut-off parameter in the
hadronic form factors.  It is apparent from fig.~2 that the reducing
effect on the total ($\gamma$,pn) cross sections due to heavier meson
exchange can be accomplished in a calculation with only pionic degrees
of freedom, provided that  an effective cut-off parameter of 800 MeV
(dot-dashed curve) is adopted.

We have investigated the usefulness of this effective approach with a
modified cut-off parameter in more detail.  To this end, we plotted in
fig.~3a the fivefold differential cross section ($d^5 \sigma^{lab} /
{d\Omega_p d\Omega_n dp_p }$) at E$_{\gamma}$=100~MeV and T$_p$=40~MeV.
In order to make the discussion more transparant only $\pi$ currents
have been included. The proton and neutron angles are determined
relative to the direction of the photon momentum. Immediately, we remark
the dominance of  back-to-back emission in the two-nucleon emission
process. The driving mechanism in this back-to-back dominance is the
missing momentum $\bf{P} = \bf{p}_a + \bf{p}_b - \bf{q}_{\gamma}$,
with $\bf{p}_{a}$ ($\bf{p}_{b}$) the momentum of the first (second)
outgoing particle and $\bf{q}_{\gamma}$ the photon momentum
\cite{Got58}. The missing momentum dependence reflects itself
in the nuclear structure factor $F_{hh'}(\bf{P})$. This is illustrated
in fig.~3b where we plotted  $F_{hh'}$(P) for the same kinematical
conditions as in fig.~3a.  It is obvious that the general structure of
the angular cross sections is a mere reflection of the F(P) dependency.
With the eye on future comparisons with the data, the differential cross
sections can best be studied under kinematical conditions which maximize
the strength.  For that purpose we have investigated angular cross
sections as a function of the outgoing proton angle in which the
corresponding neutron angle is obtained through maximizing
$F_{hh'}(\bf{P})$. This procedure corresponds with exploring the ridge
of the angular cross sections of the type displayed in fig.~3a. By doing
this the missing momentum P can be more or less kept constant over the
covered proton angle range and the large functional dependence  of the
angular cross section on the function F(P) can be ruled out. Therefore,
calculations performed along these lines are most sensitive to the
dynamics of the photoabsorption mechanism which is contained in the
factor S$_{fi}$. We have adopted  this approach to look for measurable
signs of $\rho$ absorption in the angular cross sections. In fig.~4 we
show the calculated angular cross $^{16}$O($\gamma$,pn) cross section at
two different values of the photon energy (E$_{\gamma}$ = 70 and 140
MeV).  These cross sections have been obtained through the maximizing
procedure explained earlier. We have displayed the angular cross
sections for the $\pi$ and $\pi+\rho$ absorption diagrams.  In the
calculation which involves both the $\pi$ and the $\rho$ components the
cut-off parameters of the Bonn potential have been used.  On the other
hand, the cross section  including only the $\pi$ absorption diagrams
has been obtained with the effective cut-off parameter (i.e.
$\Lambda_\pi$ = 800 MeV instead of 1200 MeV). It was observed before
that a reduction of the cut-off parameter would simulate rather well the
interference effect taking place when including $\rho$ meson exchange in
the reaction process, at least for the total integrated cross section.
This ``effective'' procedure is obviously less satisfactory in the
reproduction of the angular cross sections. At low photon energies
($E_{\gamma}$ = 70 MeV) an almost similar behaviour of both approaches
is observed. This similarity disappears with increasing photon energies.
{}From fig.~4 it is clear that as the $\rho$ meson absorption gains in
importance, the angular cross section evolves to an almost flat course,
an effect which cannot be reached when only pionic degrees are accounted
for.  This effect will be even more pronounced at higher photon
energies. We conclude that although the ``effective'' approach is
satisfactory for the integrated ($\gamma$,pn) cross section, it looses
its usefulness once we turn to angular cross sections at higher photon
energies. As such a systematic study of the angular cross section at a
fixed value of the missing momentum P,  seems to be a better tool than
the integrated cross section to understand the underlying physics of the
($\gamma$,NN) reactions. Experimental efforts to actually measure the
angular cross sections are in progress \cite{McGregor}.

As could be expected, with increasing photon energies the $\Delta$
isobar components gain in importance.  This has been illustrated in
fig.~5a in which we compare the mesonic ($\pi + \rho$) contribution to
the ($\gamma$,pn) cross section with the full calculation, which apart
from the $\pi$ and the $\rho$  has a $\Delta$(1232) part.  As has been
explained in section 2, we adopt an energy dependent nonrelativistic
$\Delta$ propagator and account for a decay width. For photon energies
exceeding 150 MeV, the $\Delta$ current is the dominant contribution to
the ($\gamma$,pn) cross section. In fig.~5 we have also plotted the
results obtained in the commonly adopted static $\Delta$ approximation
(neglecting the energy dependence and the decay width in the $\Delta$
propagator). It is clear that for pn emission  the static $\Delta$
approximation is not able to reproduce the $\Delta$ resonance peak and
gives rise to results that differ quite drastically from the energy
dependent results.  Surprisingly, this seems to be even the case at
relatively low photon energies.

Let us now turn to the ($\gamma$,pp) reaction.  In a direct knockout
model, we expect the $\Delta$ current to dominate the ($\gamma$,pp)
cross section as the major meson exchange currents only contribute when
a photon gets absorbed by a proton-neutron pair. We have found that the
mere influence of the mesonic currents, which is restricted to $\rho$,
$\sigma$ and $\omega$ absorption through the pair diagram (fig.1c), on
the pp cross sections  is a slightly modifying effect at photon energies
below 100 MeV where the $\Delta$ component results in very small cross
section. In this energy range the ($\gamma$,pp) cross section is
negligible in comparison with the ($\gamma$,pn) contribution. In fig.~5b
we display the $\Delta$ contribution to the pp channel for the two types
of $\bigtriangleup$ propagators discussed above. Unlike the
proton-neutron emission reaction, the two proton emission reaction is
rather well described by the static $\bigtriangleup$ propagator for
photon energies below 180 MeV. This different behaviour in the
($\gamma$,pn) and the ($\gamma$,pp) channel can be attributed to the
destructive interference between the direct and the exchange matrix
elements.

In order to get some handle on the realistic character of our
calculations, we have calculated the contribution of the ($\gamma$,pn)
and the ($\gamma$,pp) to the total photoabsorption strength in
$^{16}$O. This procedure involves, apart from the sum over all occupied
single-particle states, an integration over the solid angles and the
kinetic energies of both escaping particles.   In the peak of the
$\bigtriangleup$(1232), the  ($\gamma$,pn) contribution to the total
photoabsorption strength is predicted to be 2.40 mb. The total
photoabsorption cross section in the $\Delta$ peak is measured to be
about 7 mb for $^{16}O$ \cite{photo}.  The major contribution of this
strength will go into $\pi$ decay of the $\bigtriangleup$.  In this
sense, the predicted 2.40 mb appears like an overestimation.   It has to
be kept in mind that our results have been obtained in the factorized
Gottfried approach which does not involve any type of FSI. The latter is
known to yield a reduction of the cross section \cite{Giu92,new93}. On
top of that, in a previous work we have shown that the Gottfried
approximation will generally lead to an overestimation of the full
unfactorized cross section \cite{Ryc92}. In passing it is worth
mentioning that we predict a value of 5.26 mb for the contribution of the
pn channel to the $\Delta$ photoabsorption peak when we discard the
$\rho$ contribution to the $\Delta$ diagrams of fig.~1(d). This should
be considered as a totally unrealistic value.  Inclusion of $\rho$
degrees of freedom in the construction of the $\Delta$ currents turns
out to be indispensable in order to arrive at a realistic description of
the expected strength in the ($\gamma$,pn) channel.

It was already noted by Riska \cite{riska2} - in calculations
with a static $\Delta$ current - that the main effect of the $\Delta$ current
with $\rho$ exchange is to strongly reduce the effect of the $\Delta$ current
with $\pi$ exchange. We confirm here  - for a $\Delta$ current with
an energy-dependent $\Delta$ propagator -  the importance of retaining both
$\pi$ and $\rho$ contributions.

\section{Summary}

We have investigated the $\pi$, $\rho$ $\sigma$, $\omega$
and $\Delta$ contributions to the
($\gamma$,NN) cross section at intermediate energies.
In order to keep the physics of the photoabsorption separate from
the nuclear-structure effects we have used the factorized Gottfried approach.
The $\sigma$ and $\omega$ meson were checked not to  give any
appreciable contribution to the cross sections. On the other hand, the
$\pi$-$\rho$ interference term yields a considerable reduction of  the
leading $\pi$ exchange current contribution. We have shown that the
influence of $\rho$ meson absorption on the total cross section can be
simulated quite well in a  calculation which involves only $\pi$ degrees
of freedom when introducing an effective
cut-off mass of 800 MeV in the hadronic $\pi$NN form factor.
This statement, however,
does no longer hold for the
angular cross sections at a fixed value of the missing momentum P,
in which the effect of $\rho$ absorption is clearly reflected in
the shape and where it is consequently important to include both $\pi$
and $\rho$ meson degrees of freedom.  This points towards a selective type of
angular cross sections being a
better research tool than integrated ($\gamma$,NN) strength
distributions to pin down the physics of photon induced two-nucleon
emission processes. We have also assessed the role of the $\Delta$ current
which has been observed to dominate the MEC once the photon energy exceeds
150~MeV. It was found important to retain both the pion and rho
contributions to the $\Delta$ current because of the strong destructive
interference between them.  Including solely the $\pi$ decay of the
$\Delta$ isobar leads to a severe overestimation of the strength in the
resonance region.
The static approach of the $\Delta$ current
seems only to be reasonable for the ($\gamma$,pp) reaction and this up to
photon energies of 180 MeV.  Summarizing, we have observed a general
sensitivity of the calculated ($\gamma$,NN) cross sections to effects
related to $\rho$ degrees of freedom.  Accordingly, extensive
($\gamma$,NN) measurements might help elucidate the short-range features
of the NN interaction.
\\
{\em Acknowledgement}
\\
This work has been supported by the National Fund for Scientific
Research (NFWO), the Institute for Scientific Research in Industry and
Agriculture (IWONL) and in part by the NATO through the research grant
NATO-CRG920171.

\newpage

\begin{figure}
\caption{The Feynman diagrams used to calculate the nuclear current
operator.  Diagrams (a-c) correspond to the MEC  and diagram (d)
to the $\bigtriangleup$ isobar.}
\end{figure}

\begin{figure}
\caption{Different mesonic absorption mechanisms contributing to the
$^{16}$O($\gamma$,pn) cross section in planar kinematics as a function
of the photon energy. The contributions from the $\pi$-MEC (dashed
curve-$\Lambda_\pi$ = 1200 MeV), $\rho$-MEC (dotted curve-$\Lambda_\rho$
= 1500 MeV) and $\pi$$\rho$-MEC (full curve-$\Lambda_\pi$ = 1200 MeV,
$\Lambda_\rho$ = 1500 MeV) are shown. Also shown is the $\pi$-MEC
contribution with $\Lambda_{\pi}$ = 800~MeV (dot-dashed curve).}
\end{figure}

\begin{figure}
\caption{(a) Differential cross section ($d^5 \sigma^{lab} / {d\Omega_p
d\Omega_n dp_p }$) for the $^{16}$O($\gamma$,pn) reaction at $E_\gamma$ =
100 MeV and $T_{p}$ = 40 MeV. Only pionic absorption mechanisms are
accounted for. (b)
The function F($\vec{P} = \vec{p}_p + \vec{p}_n -
\vec{q}_{\gamma}$) for the kinematical conditions of (a). }
\end{figure}

\begin{figure}
\caption{The contributions to the cross section in the kinematics as
explained in the text. for $\pi\rho$-MEC and the effective $\pi$-MEC at
different photon energies in function of the proton angle. At $E_\gamma$
= 70 MeV we plotted the $\pi\rho$-MEC (full curve) and effective
$\pi$-MEC (dotted) contribution. Idem dito at $E_\gamma$ = 140 MeV
(dashed) (dot-dashed).}
\end{figure}

\begin{figure}
\caption{The $^{16}$O($\gamma$,NN) cross section as function of
the photon energy for different mesonic and isobaric absorption
mechanisms. (a) pn emission : $\pi\rho$-MEC contribution
(dashed curve) and MEC-$\bigtriangleup$ contribution as calculated with an
energy-dependent propagator (solid curve) and in the static limit (dotted
curve). (b) pp emission : curves as in (a).}
\end{figure}

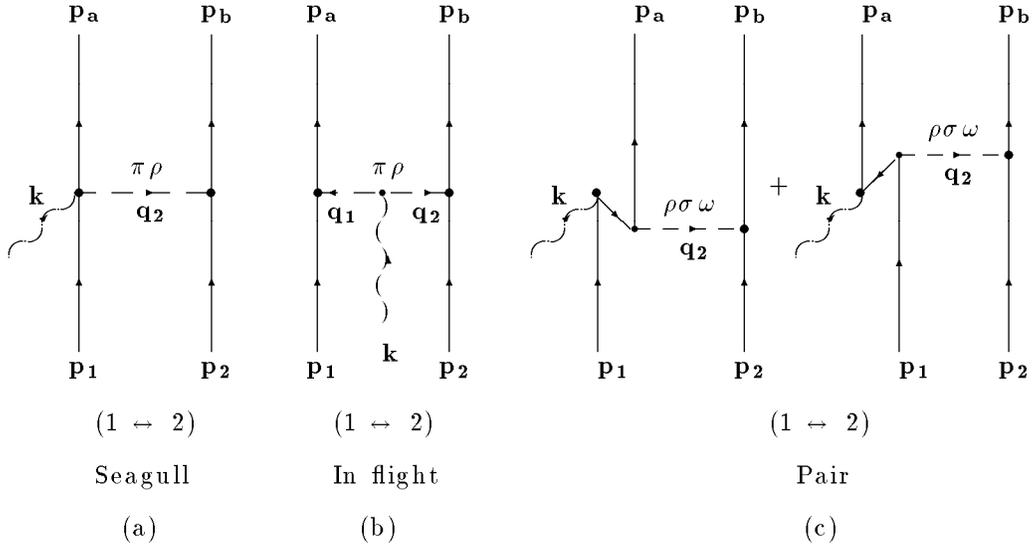
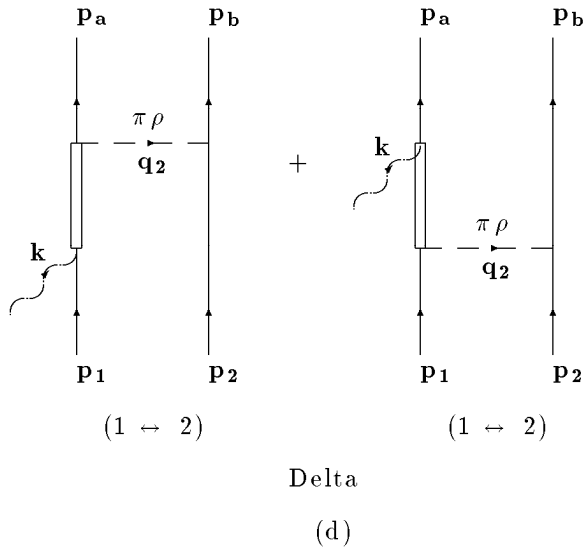
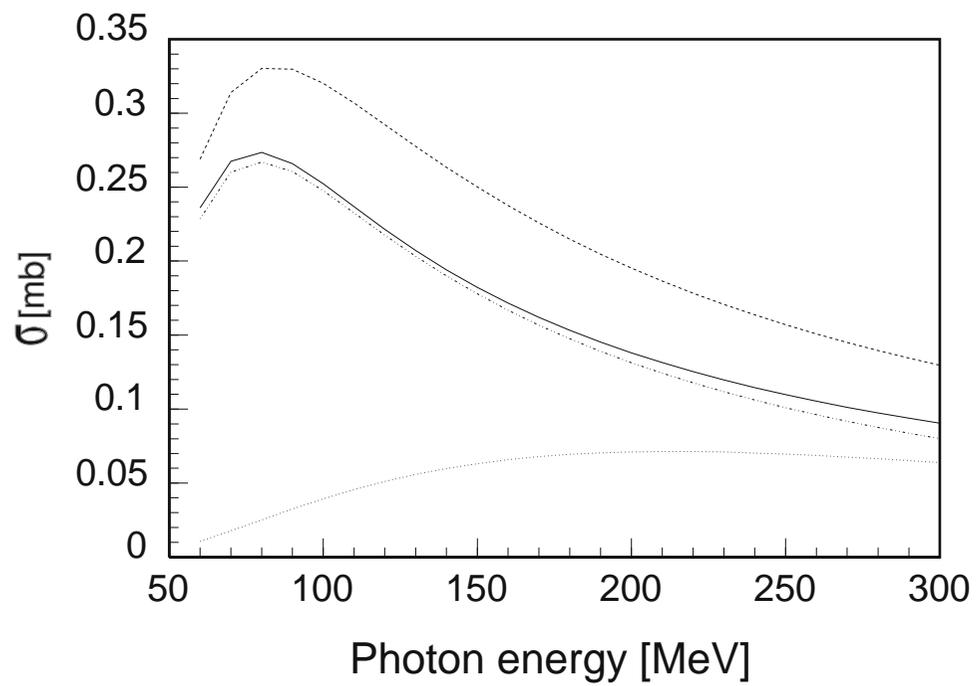
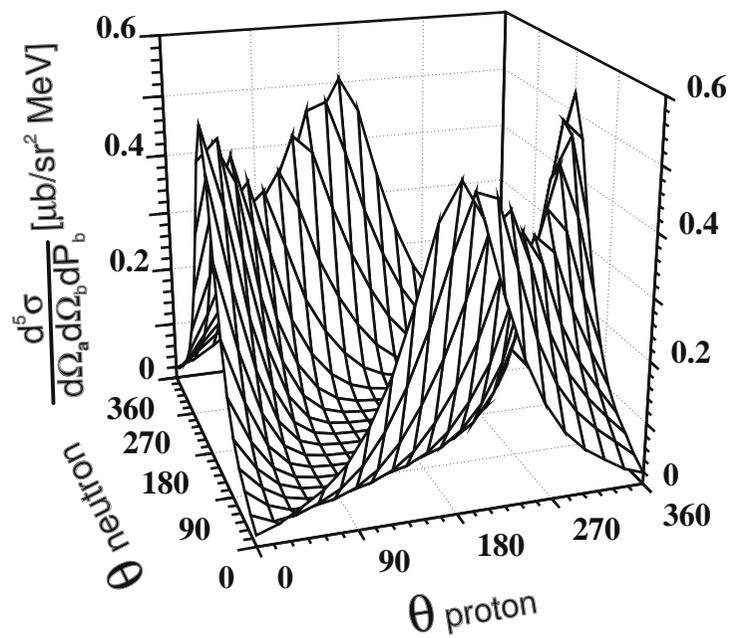
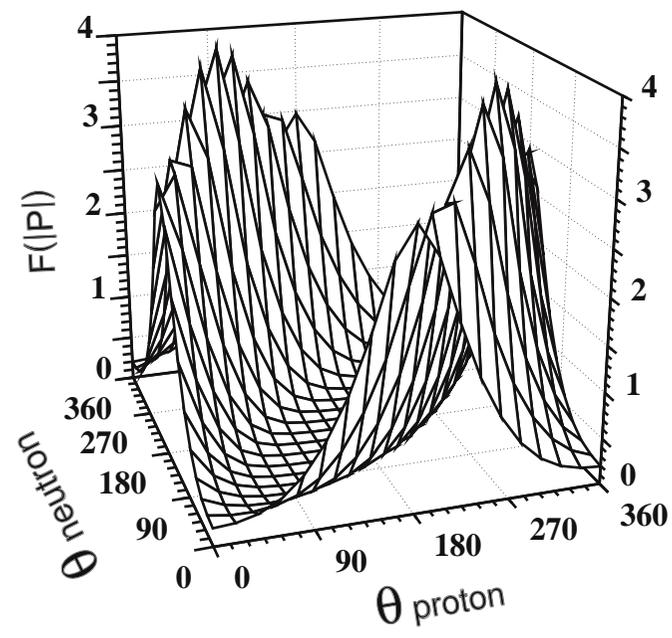
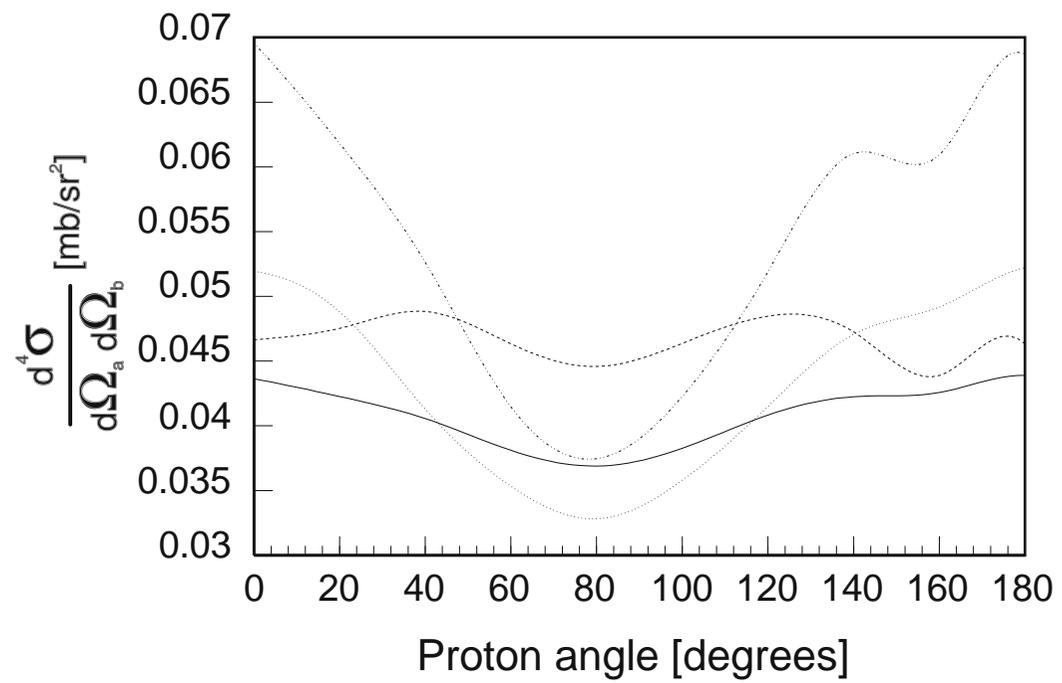
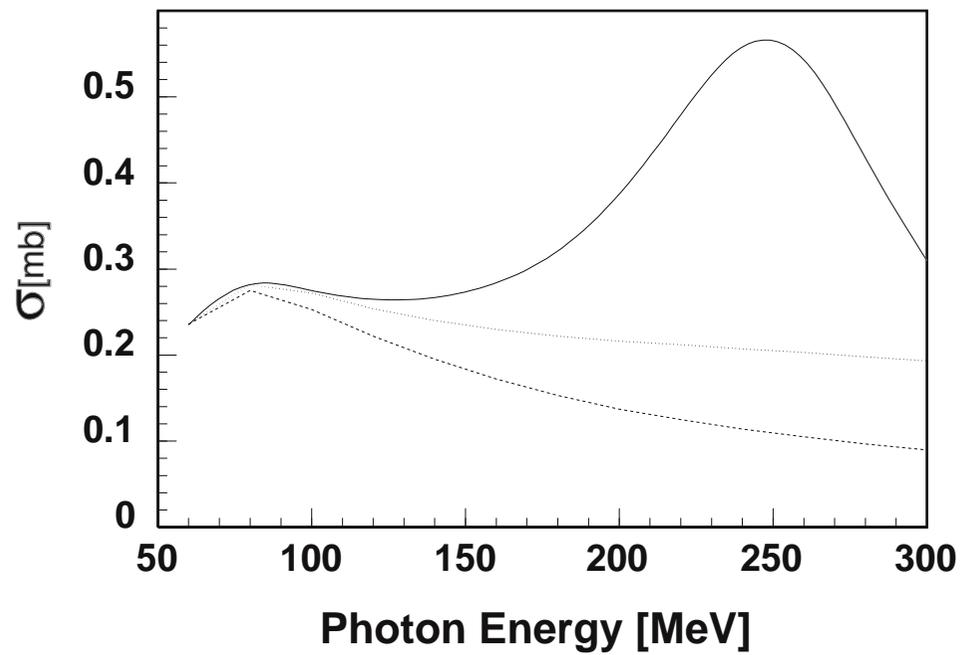
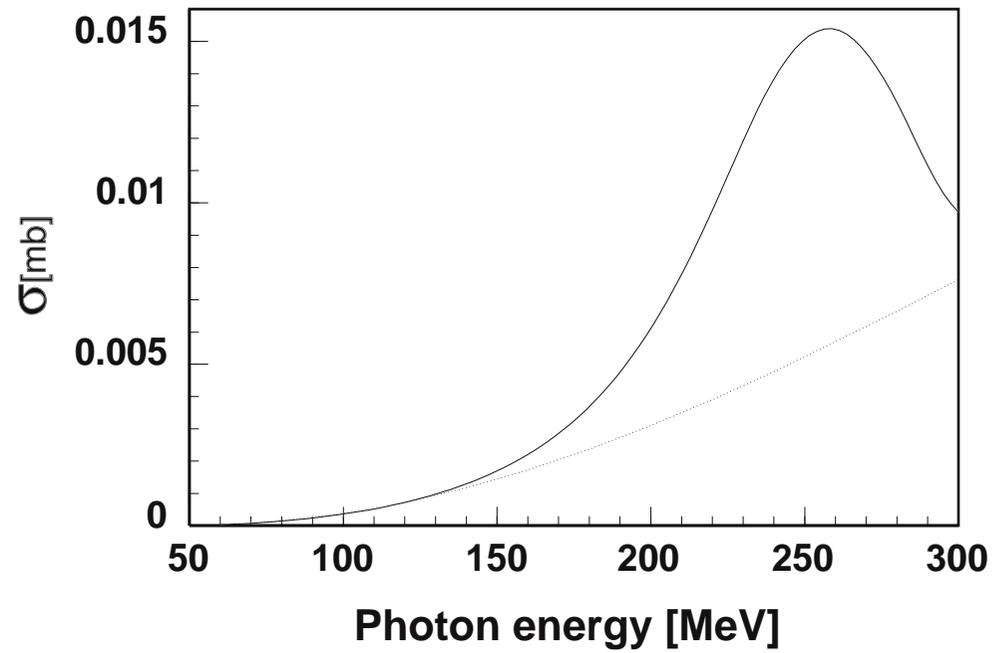
\end{document}